\documentclass[aps,prl,twocolumn,superscriptaddress,10pt]{revtex4-2}

\usepackage{graphicx}
\usepackage{dcolumn}
\usepackage{bm}
\usepackage{amsmath,amsfonts,mathtools}
\usepackage{booktabs}
\usepackage{float}
\usepackage{comment}
\usepackage[svgnames,table]{xcolor}
\usepackage{siunitx}
\usepackage{ragged2e}
\usepackage[colorlinks=true,citecolor=SteelBlue,filecolor=LimeGreen,linkcolor=SlateBlue,urlcolor=MediumPurple]{hyperref}

\DeclareSIUnit\sqrthz{\ensuremath{\sqrt{\text{Hz}}}}
\DeclareSIUnit[per-mode = symbol]\Hzasd{\Hz\per\sqrthz}
\DeclareSIUnit[per-mode = symbol]\masd{\m\per\sqrthz}
\DeclareSIUnit[per-mode = symbol]\pmasd{\pico\m\per\sqrthz}

\makeatletter
\let\cat@comma@active\@empty
\makeatother

\begin{document}

\preprint{APS/123-QED}

\title{Torsional-X Seismometer for Lunar Decihertz Gravitational-Wave Detection}

\author{Yulin Xia}
\affiliation{Department of Physics, Tsinghua University, Beijing 100084, China}

\author{Denis Martynov}
\affiliation{Institute for Gravitational Wave Astronomy, School of Physics and Astronomy, University of Birmingham, Birmingham B15 2TT, United Kingdom}

\author{Huan Yang}
\affiliation{Department of Astronomy, Tsinghua University, Beijing 100084, China}

\author{Haixing Miao}
\email{haixing@tsinghua.edu.cn}
\affiliation{Department of Physics, Tsinghua University, Beijing 100084, China}

\date{\today}

\begin{abstract}

The lunar gravitational-wave antenna concept uses the Moon as a resonant detector instrumented with precision seismometers, targeting the decihertz band between ground- and space-based observatories. We propose a compact monolithic fused-silica torsional-X seismometer that re-engineers garden-gate acceleration-to-rotation transduction for this regime through a high-tension dual-fiber suspension. Its designed millihertz-scale resonance and ultra-low mechanical dissipation enable a nearly order-of-magnitude improvement around \(0.1\,{\rm Hz}\) compared with existing lunar seismometer concepts. Achieving this performance requires room-temperature operation, where fused-silica exhibits low mechanical loss, together with subdominant actuation noise. We demonstrate a room-temperature vacuum prototype validating the operating principle and core mechanical design, and derive requirements for a future lunar implementation capable of approaching the target sensitivity.
\end{abstract}

\maketitle

\textit{Introduction.---} Ground-based and planned space-based gravitational-wave detectors probe complementary frequency bands. Current ground-based observatories, including LIGO, Virgo, and KAGRA, together with planned third-generation detectors such as the Einstein Telescope and Cosmic Explorer, are most sensitive in the $\sim 10$--$10^{4}\,\mathrm{Hz}$ band~\cite{LIGO,Virgo,KAGRA,Aplus,ET,CE}. Planned space-based missions, including LISA, TianQin, and Taiji, target the lower-frequency band of $\sim 10^{-4}$--$10^{-1}\,\mathrm{Hz}$~\cite{LISA,Tianqin,Taiji}. Between these two regimes lies the relatively unexplored decihertz window, $\sim 0.1$--$10\,\mathrm{Hz}$, which is expected to host several astrophysically important sources~\cite{DHZ_GW_source}. Access to this band would enable observations of intermediate-mass black-hole binaries~\cite{IMBHs} and, in combination with ground-based detectors, provide earlier tracking and improved sky localization of merging stellar-mass binaries.

These scientific opportunities have motivated a range of decihertz detector concepts~\cite{,Lunar_GW_Paik,TOBA,DECIGO,MAGIS,LGWA,Torpedo}. Among them, the Lunar Gravitational-Wave Antenna (LGWA) exploits the Moon's low ambient seismic background and its response to gravitational waves to target the decihertz band~\cite{Lunar_GW,Lunar_GW_Chen}. In this approach, the Moon acts as a resonant-bar detector instrumented with high-precision seismometers that measure the minute ground motions induced by passing gravitational waves~\cite{Lunar_response_Dyson,Lunar_responce_Jan,Lunar_responce_Yan,Lunar_responce_Yan_2D,Lunar_responce_Nodtvedt,Lunar_responce_Lognonne}. Early lunar seismometer concepts consider both optomechanical and cryomagnetic devices, such as the LGWA-opto (-cryo)~\cite{LGWA}, and Beijing Normal University (BNU) designs~\cite{BNU,BNU_erratum}. Recent optomechanical designs explore Watt-linkage~\cite{LGWA_Watt-linkage1,LGWA_Watt-linkage2} with niobium or silicon test masses to achieve sub-hertz resonances and material-limited quality factors of $10^4$--$10^6$~\cite{LGWA_reviews,LGWA_missions}.

Other approaches to sub-hertz seismometers include LaCoste zero-length springs~\cite{LaCoste_spring}, spring--antispring systems~\cite{spring_antispring}, leaf-spring designs~\cite{STS,leaf_spring}, and garden-gate configurations~\cite{garden_gate_fiber,garden_gate}. These concepts reduce the residual restoring stiffness using engineered springs or flexures. For lunar gravitational-wave detection, however, the required parameter regime is more restrictive than a direct extrapolation of conventional designs. Suppressing suspension thermal noise in the decihertz band requires pushing the relevant mechanical resonance to the ultra-low frequency while retaining a compact and low-loss instrument. This makes the problem a coupled optimization of geometry and material choice, rather than simply a softer spring or lower-frequency pivot~\cite{Low_frequency_resonator_creep}.

Here we propose a compact monolithic fused-silica torsional-X (TX) seismometer that re-engineers the garden-gate principle for the lunar-GW regime. A center-of-mass offset converts horizontal acceleration into torsional motion, while a high-tension dual-fiber suspension compensates the resulting static gravitational torque and stiffens the unwanted translational modes out of the decihertz band. This architecture realizes a few-millihertz torsional resonance in a tens-of-centimeters-scale payload, exploiting a central advantage of torsional inertial sensors~\cite{Large6D,Metal6D,Compact6D,Torpedo,TOBA,UF_Torsional,Ground_test_LISA}. The use of fused-silica is equally essential: its low dissipation, nearly ideal elasticity, and negligible creep have been central to reducing suspension thermal noise in gravitational-wave detectors~\cite{Fused_silica_fiber_dissipation,Fused_silica_loss_in_LIGO,Suspension_thermal_noise}. Together with shot-noise-limited interferometric readout~\cite{LIGO} and femtonewton-level electrostatic actuation~\cite{LISA_actuator}, the TX design is projected to improve the strain sensitivity of existing LGWA seismometer concepts by nearly an order of magnitude around \(0.1\,{\rm Hz}\). Beyond the common challenges of lunar deployment, this approach requires maintaining the suspension in a controlled warm, low-loss regime, typically near room temperature, rather than passively operating at the cryogenic temperature of permanently shadowed regions, since fused-silica exhibits a broad cryogenic dissipation peak~\cite{Fused_silica_loss_at_cryo_temp}. We validate the operating principle with a room-temperature vacuum prototype, confirming the core optomechanical dynamics and quantifying the remaining subsystem requirements for a realistic lunar instrument.

\begin{figure}[t]
    \centering
        \centering
        \includegraphics[width=\linewidth]{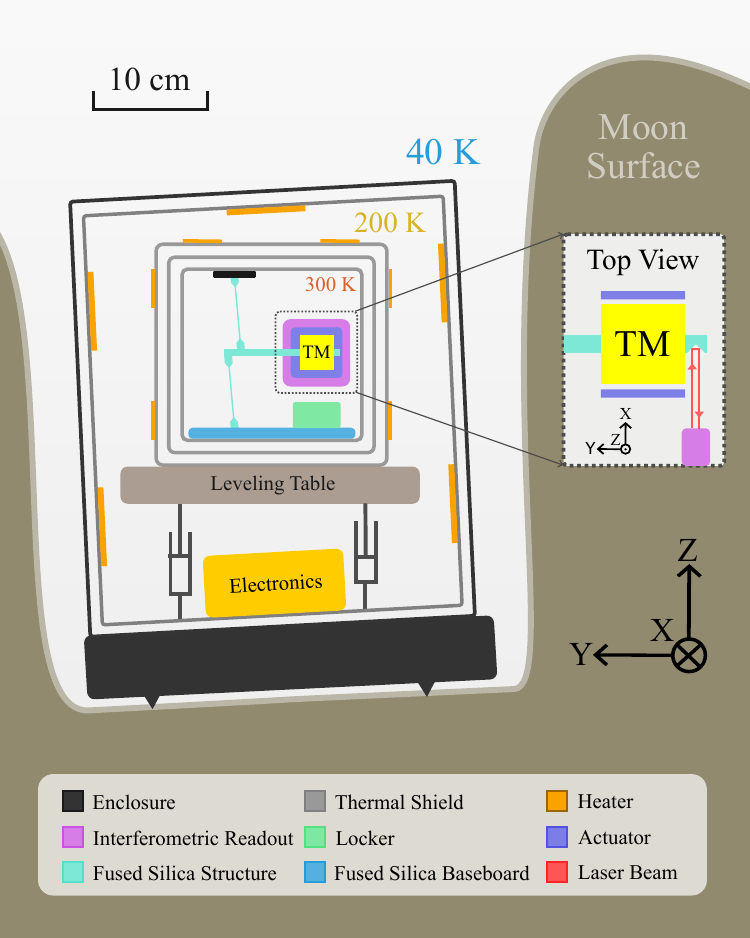}
        \caption{\justifying Conceptual design of the lunar seismometer station. The TX payload is housed in a thermally controlled station and mounted on an internal leveling platform. The unit shown senses motion along one horizontal degree of freedom; a full horizontal implementation uses two identical, orthogonally oriented units to measure both \(x\) and \(y\) axes.}
        \label{fig:struct_design}
\end{figure}

\begin{figure}[t]
    \centering
        \centering
        \includegraphics[width=\linewidth]{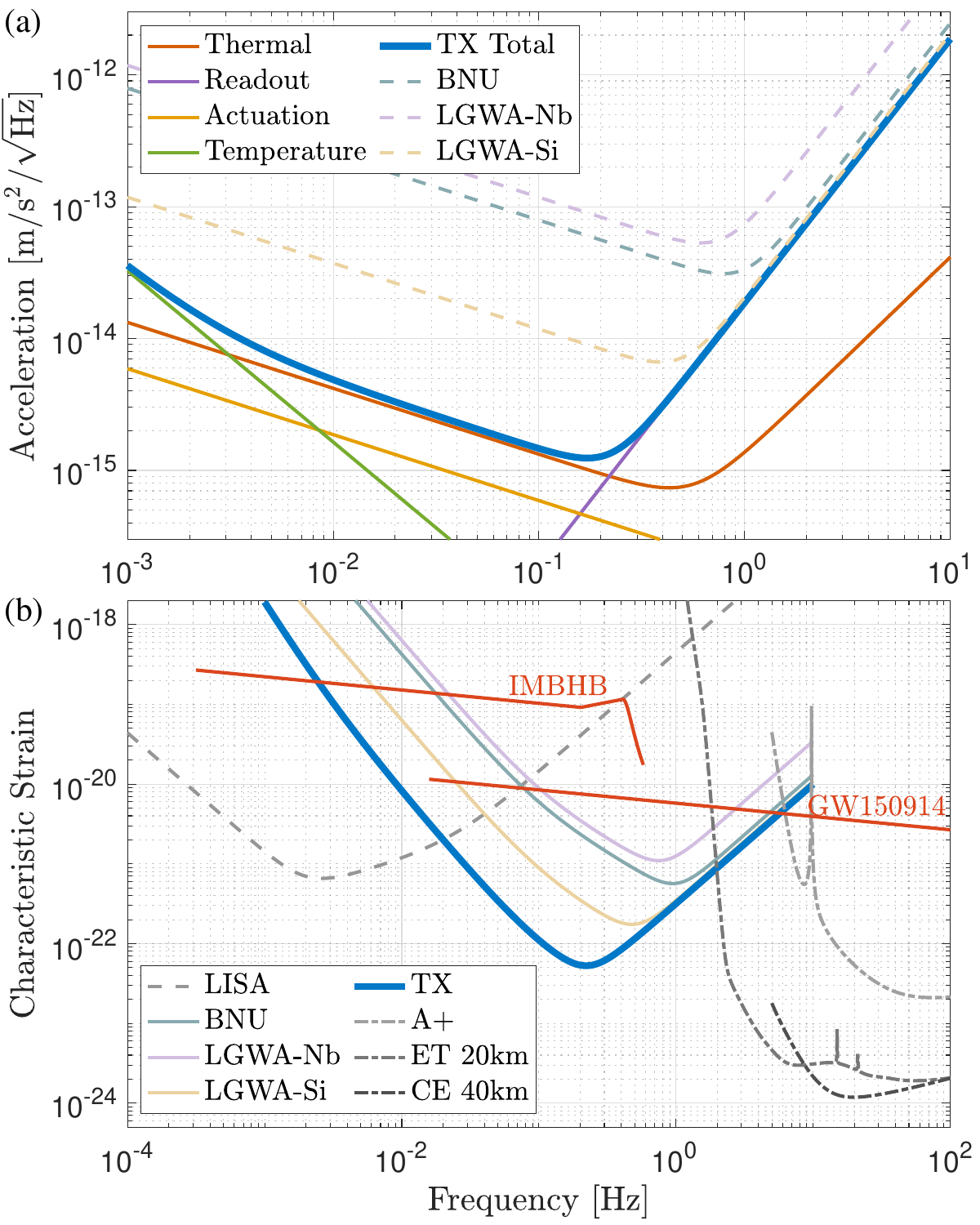}
        \caption{\justifying
                Sensitivity of the TX concept.
                (a) Acceleration-noise budget of the TX seismometer, compared with the BNU and LGWA concepts. The TX thermal, readout, actuation, and temperature-induced noise curves follow Eqs.~\eqref{eq:thermal_noise}, \eqref{eq:readout_noise}, \eqref{eq:actuation_noise}, and \eqref{eq:temp_noise}, respectively. 
                (b) Characteristic gravitational-wave strain sensitivity, compared with LISA, the BNU design, LGWA concepts, Advanced LIGO Plus (A+), the Einstein Telescope (ET), Cosmic Explorer (CE), representative intermediate-mass black-hole binary signals, and GW150914.
                }
        \label{fig:sensitivity}
\end{figure}

\textit{Design concept and science goal.---}
Figure~\ref{fig:struct_design} illustrates the conceptual design of the TX seismometer station. The instrument is envisioned for deployment in a pre-leveled pit within a permanently shadowed lunar region. The station comprises the core TX seismometer and three supporting subsystems: a temperature-control system with multiple thermal shields and heaters, an internal leveling table, and a locking mechanism for launch, landing, and deployment.

The core payload is a monolithic fused-silica frame suspended from above and below by two laser-pulled fibers, forming a torsional pendulum. For the test mass, we choose tungsten and it allows us to achieve required moment of inertia within a compact size. The center of mass is intensionally offset from the rotation axis by $l_c$ to  efficiently convert the horizontal ground acceleration, $a_g$, into yaw motion, $\theta_z$, which is measured by a laser interferometer.

The same offset also introduces a static gravitational torque, which must be compensated by the suspension geometry. Unlike a conventional torsional pendulum, the TX design uses two suspension fibers symmetrically tilted by an angle $\alpha$. The corresponding displacement of the attachment points allows the fiber tension to generate a static counter-torque. Using the key design parameters summarized in Table~\ref{tab:TX_parameters}, the torsional eigenfrequency is $f_{\theta}=4.6\,\mathrm{mHz}$ due to the weak torsional stiffness of the suspension, while the horizontal pendulum mode is raised to $f_{x}=4.0\,\mathrm{Hz}$ by the large tensile stress in the fibers~\cite{Glasgow_fiber_strength}. 
For the baseline noise budget, we adopt an effective fused-silica loss angle of \(\phi=10^{-7}\), which accounts for both bulk and surface dissipation and is consistent with demonstrated suspensions~\cite{Fused_silica_fiber_dissipation, Fused_silica_loss_in_LIGO}.

The dominant noise sources limiting the seismometer sensitivity are mechanical dissipation, which gives rise to thermal noise, and quantum fluctuations of the optical field in the laser interferometer, which set the shot-noise limit of the readout. 
For an optical readout with lever arm \(l_r\), the readout response, including the equivalent thermal-acceleration contributions, is derived in Supplemental Material and can be written as
\begin{equation}
\label{eq:xrot}
    x_{\rm ro}(f)
    \simeq
    -\frac{a_g+l_c a_\theta^{\rm th}}
    {4\pi^2 \left[f^2-f_{\theta}^2(1+i\phi)\right]} \frac{l_r}{l_c}
    - \frac{a_g+\epsilon a_x^{\rm th}}{4\pi^2f_x^2}.
\end{equation}
Here \(\epsilon=(l_r+l_c)/l_c\) is the readout-geometry factor for the horizontal thermal contribution. 
The term \(a_\theta^{\rm th}\) denotes the equivalent angular acceleration noise associated with torsional-mode dissipation, while the term \(a_x^{\rm th}\) denotes the horizontal-mode thermal acceleration noise.

\begin{table}[t]
    \centering
    \caption{Key parameters of the TX seismometer}
    \label{tab:TX_parameters}
    \resizebox{\linewidth}{!}{
    \begin{tabular}{lll}
    \toprule
    Parameter & Description & Value \\
    \midrule
    \multicolumn{3}{c}{\textit{Mechanical parameters}} \\
    \midrule
    $g_{\rm moon}$ & Lunar gravitational acceleration & $1.6\,\mathrm{m/s^{2}}$ \\
    $V$ & Test mass volume & $4\times4\times4\,\mathrm{cm^3}$ \\
    $m$ & Test mass mass & $1.2\,\mathrm{kg}$ \\
    $l_c$ & COM offset from rotation axis & $10\,\mathrm{cm}$ \\
    $r$ & Fiber radius & $50\,\mu\mathrm{m}$ \\
    $L$ & Fiber length & $6.5\,\mathrm{cm}$ \\
    $\alpha$ & Fiber tilt angle & $2.4^\circ$ \\
    $\sigma$ & Tensile stress in fiber & $3\,\mathrm{GPa}$ \\
    $\phi$ & Mechanical loss angle & $10^{-7}$ \\
    $T$ & Operating temperature & $300\,\mathrm{K}$ \\
    $f_{\theta}$ & Torsional eigenfrequency & $4.6\,\mathrm{mHz}$ \\
    $f_{x}$ & Horizontal eigenfrequency & $4.0\,\mathrm{Hz}$ \\
    \midrule
    \multicolumn{3}{c}{\textit{Readout and control parameters}} \\
    \midrule
    $\lambda$ & Laser wavelength & $1064\,\mathrm{nm}$ \\
    $P$ & Laser power & $20\,\mathrm{mW}$ \\
    $l_r$ & Readout lever arm & $12\,\mathrm{cm}$ \\
    $U_0$ & Maximum control voltage & $20\,\mathrm{V}$ \\
    $d_0$ & Electrode gap & $2\,\mathrm{mm}$ \\
    $A_e$ & Electrode area & $4\times4\,\mathrm{cm^2}$ \\
    \bottomrule
    \end{tabular}}
\end{table}

The total acceleration-referred thermal-noise spectral density is therefore the sum of the direct torsional contribution and the horizontal-mode contribution projected onto the same optical readout. 
Using the fluctuation--dissipation theorem~\cite{Fluctuation-disspation}, we obtain
\begin{equation}
S_a^{\rm th}(f)
=
\frac{8 \pi k_B T}{m}
\left[
\frac{f_{\theta}^2}{f}
+
\left(\frac{l_r+l_c}{l_r}\right)^2
\frac{f^3}{f_{x}^2}\eta
\right]\phi\,,
\label{eq:thermal_noise}
\end{equation}
where \(\eta\) is the horizontal-mode dilution factor. This dilution arises because the horizontal restoring stiffness is tensile-dominated, while dissipation is localized mainly in boundary bending~\cite{Damping_dilution}. In our design, \(\eta\simeq10^{-3}\), which strongly suppresses the horizontal thermal-noise contribution to the readout.
For the interferometric readout noise, we assume the same shot-noise-limited displacement sensitivity as in Refs.~\cite{LGWA,LGWA_reviews,LGWA_missions}. The corresponding acceleration-referred noise spectrum is
\begin{equation}
S_a^{\rm ro}(f)
=
4 \pi^2
\frac{\lambda_0 h c}{P}
\frac{l_r^2}{l_c^2}
\left|f_{{\theta}}^2(1+i\phi)-f^2\right|^2\,.
\label{eq:readout_noise}
\end{equation}
The above two noise contributions define the fundamental sensitivity limit of the TX configuration and set the requirements on the supporting subsystems, as summarized in Fig.~\ref{fig:sensitivity}(a). Owing to the millihertz torsional eigenfrequency and the low dissipation of fused-silica fibers, it achieves exceptionally low acceleration noise which is around $10^{-15}\,\rm m/s^2/\sqrt{Hz}$ at $0.1\,\rm Hz$. 

To convert acceleration noise into gravitational-wave strain, we adopt the lunar response model used in the LGWA study~\cite{LGWA_missions}. The resulting absolute strain sensitivity is therefore model dependent, since the lunar response can be affected by lateral heterogeneity, topography, and location-dependent resonant amplification~\cite{Lunar_amplification_Yan}. For the baseline estimate shown in Fig.~\ref{fig:sensitivity}(b), we use a piecewise fit to the LGWA response function. The resulting characteristic sensitivity of the TX concept is compared with representative lunar seismometer concepts~\cite{LGWA_missions,BNU}, the target sensitivities of ground- and space-based detectors, and illustrative signals from an intermediate-mass black-hole binary with total mass \(10^4\,M_\odot\) at redshift \(z=1\)~\cite{IMR}, together with GW150914~\cite{GW150914}.

\begin{figure*}[t]
    \centering
    \begin{minipage}[b]{0.475\textwidth}
        \centering
        \includegraphics[width=\linewidth]{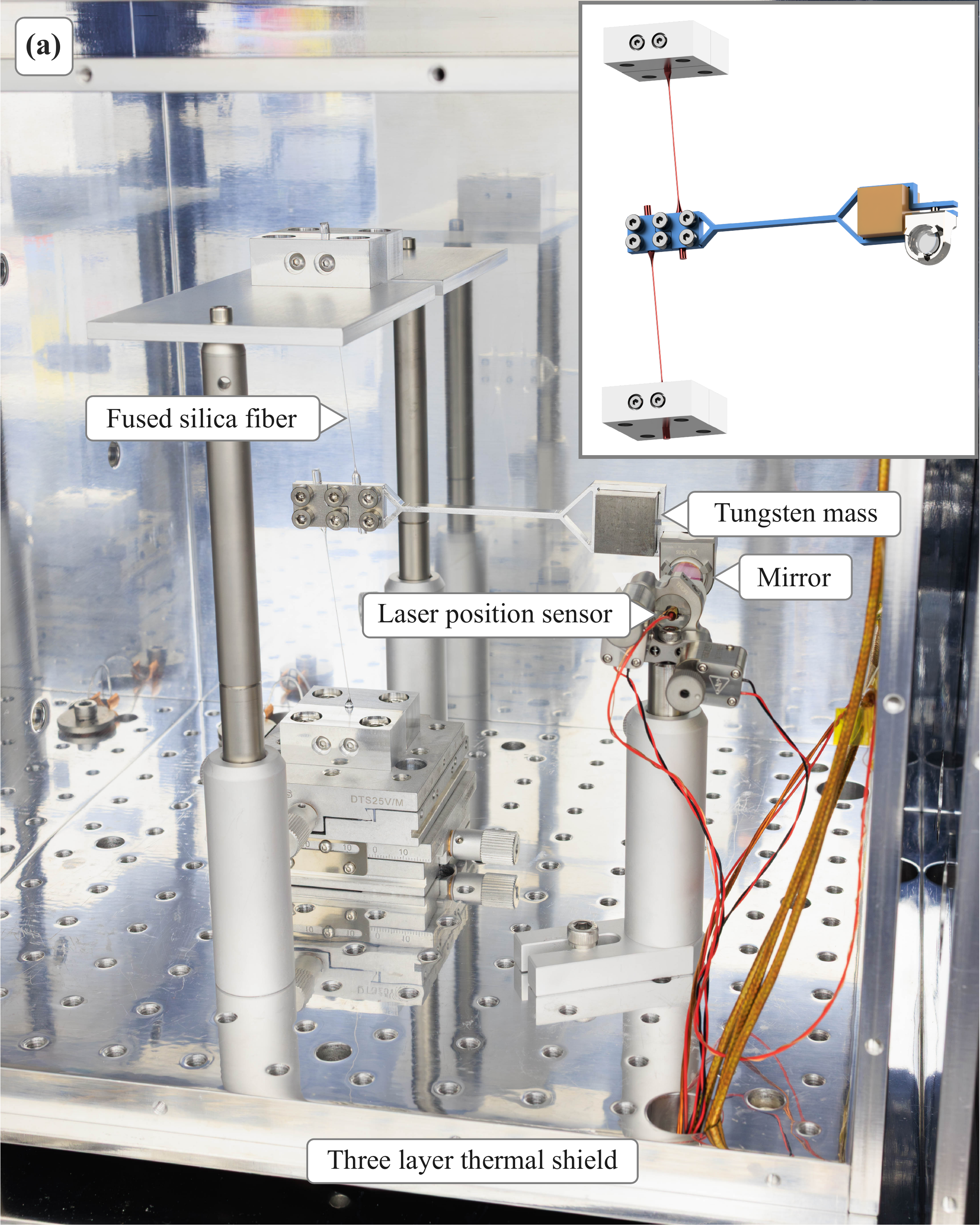}
    \end{minipage}
    \hfill
    \begin{minipage}[b]{0.475\textwidth}
        \centering
        \includegraphics[width=\linewidth]{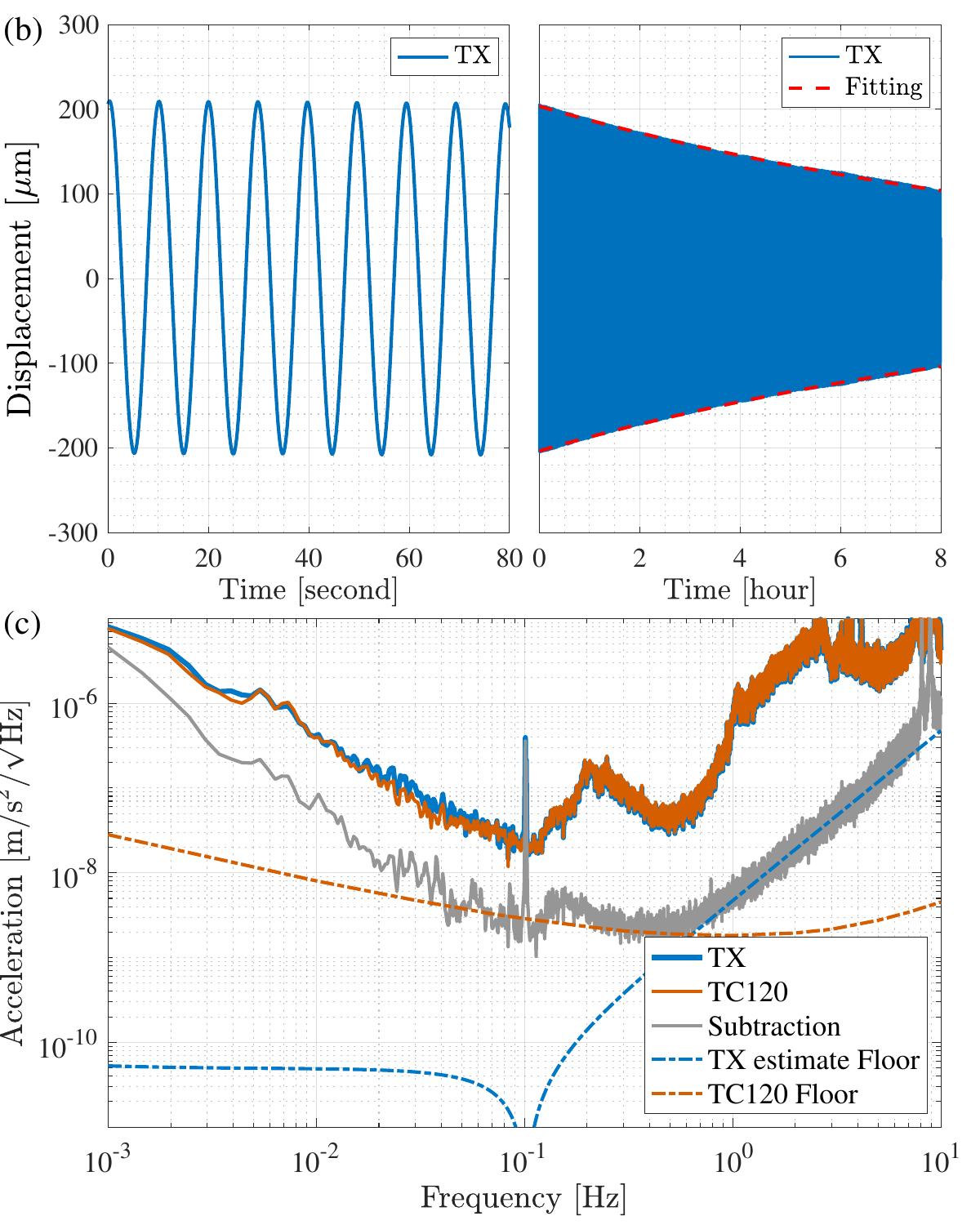}
    \end{minipage}
    \caption{\justifying
    Experimental characterization of the TX prototype.
    (a) Photograph of the prototype, showing the fused-silica fiber suspension and LPS readout. The upper right corner shows a false-color rendering of the core structure, which highlights the fused-silica fibers (red), aluminum frame (blue), and tungsten test mass (yellow).
    (b) Ring-down measurement.
    (c) Calibrated acceleration spectra measured by the TX and TC120, together with their subtraction, and relevant or estimate noise floors.
    }
    \label{fig:prototype_summary}
\end{figure*}

\textit{Proof-of-principle.---}
To verify the working principle of the proposed design, we have constructed and tested a room-temperature prototype with fused-silica suspension. Figure~\ref{fig:prototype_summary}(a) shows the experimental setup. The two fiber suspension points are vertically aligned to eliminate gravitational stiffness; this alignment can be fine-tuned using the translation stages below the suspension. The test mass motion is read out by a laser position sensor (LPS)~\cite{LPS,LPS_new}, which provides picometer-level displacement sensitivity in the decihertz band. The entire assembly is enclosed by a three-layer thermal shield under a vacuum of $10^{-6}\,\mathrm{mbar}$.

Key mechanical parameters are chosen to follow the lunar design in Table~\ref{tab:TX_parameters}, with a few modifications required by the terrestrial environment and laboratory constraints. First, to maintain a comparable offset gravitational torque and fiber tensile force under Earth's gravity, the suspended tungsten mass is reduced by a factor of six, reflecting the ratio between lunar and terrestrial gravity. Second, the laboratory-pulled fused-silica fibers have a yield stress of \(\sim 1\,\mathrm{GPa}\). To keep a safety factor of three, we set the prototype tensile stress to \(0.33\,\mathrm{GPa}\). The fiber radius is therefore increased to \(150\,\mu\mathrm{m}\) to keep the tensile force comparable to that of the lunar design. These changes shift the prototype torsional and horizontal eigenfrequencies to \(0.10\,\mathrm{Hz}\) and \(9.7\,\mathrm{Hz}\), respectively. Despite these modifications, the prototype remains a direct test of the key dynamical principle.

We characterize the prototype eigenfrequency and quality factor using a ringdown measurement. As shown in Fig.~\ref{fig:prototype_summary}(b), the test mass oscillation period is $9.9\,\mathrm{s}$, corresponding to an eigenfrequency of $0.10\,\mathrm{Hz}$, consistent with the design parameters. An $8$-hour ringdown measurement and a fit to the amplitude envelope yield $Q=1.4\times10^4$, or equivalently a loss angle $\phi=7.3\times10^{-5}$. This loss is already below that of comparable metal-based designs ($\sim10^{-4}$). The remaining gap is likely dominated by clamping loss at the current metal--silica interfaces, as discussed in the Supplemental Material. A fully monolithic fused-silica implementation would potentially reach the required effective loss angle~\cite{Fused_silica_loss_in_LIGO}.


Beyond the ringdown measurement, we have also tested the TX prototype as a horizontal seismometer by comparison with a commercial broadband instrument, the Trillium Compact Horizon 120s (TC120). After calibration using the mechanical transfer function, the amplitude spectral density recorded by the TX prototype agrees well with that of the TC120 in the decihertz band, as shown in Fig.~\ref{fig:prototype_summary}(c). The subtraction reflects the combined noise of both instruments. Above $1\,\mathrm{Hz}$, the subtraction is dominated by LPS readout noise, corresponding to a displacement noise floor of tens of $\mathrm{pm/\sqrt{Hz}}$. 
This is a limitation of the laboratory readout rather than of the TX architecture: compact optical-resonator and interferometric readouts have demonstrated sub-picometer to femtometer-level sensitivity~\cite{Carter_compact_optical_resonator,HCT_sub_femtometer}, while superconducting readouts provide an alternative high-sensitivity route for cryogenic inertial sensors~\cite{SQUID_Paik,SQUID_review}.
Between $0.03\,\mathrm{Hz}$ and $1\,\mathrm{Hz}$, the TC120 noise floor becomes the main limitation, while below $0.03\,\mathrm{Hz}$ the subtraction is likely dominated by low-frequency ground tilt, air-current and temperature-fluctuation noise, since the TC120 operates outside the vacuum chamber and thermal shield.
Taken together, these results validate the essential TX optomechanical architecture and provide an experimental foundation for the conceptual design, motivating a quantitative assessment of the remaining gap and subsystem requirements for a lunar gravitational-wave instrument.

\textit{From prototype to lunar instrument.---}
The key question is therefore no longer whether the TX architecture works in principle, but what subsystem requirements must be met to realize a complete lunar instrument. In addition to the fundamental limits from thermal and readout noise, any low-eigenfrequency, high-precision horizontal seismometer---including existing lunar concepts---need to account for tilt coupling, actuation noise, leveling control, temperature fluctuations, and magnetic coupling. For the TX design, the reduced thermal-noise floor and room-temperature operating point make these auxiliary requirements especially important, particularly for electrostatic actuation, leveling, and temperature stability. We take the thermal and readout noises as the irreducible floor and require all supporting subsystems to remain subdominant under realistic engineering assumptions.

\textit{Electrostatic control and leveling.---}
Ground tilt couples to the horizontal readout through gravity, with a displacement enhancement of order $g/\omega^2$~\cite{TTX1,TTX2}. This coupling becomes increasingly important at low frequencies, especially for soft horizontal seismometers. In addition, a high-$Q$ test mass resonator can undergo large resonant motion driven by ambient lunar excitations~\cite{Lunar_background,Lunar_seis_simu}. A practical lunar seismometer therefore cannot operate as a purely free-running sensor; it requires force balance to suppress large motions and keep the interferometric readout within its linear regime. Here we consider an electrostatic actuation scheme analogous to that used in LISA~\cite{LISA}. Two electrodes are placed symmetrically about the test mass with nominal gap $d_0$, and differential voltages $U_0/2 \pm \Delta U$ produce a net control force. For drive electronics with voltage-noise spectrum $S_V$, the resulting actuation-induced acceleration noise is
\begin{equation}
S_a^{\rm act}(f)
=
\frac{\epsilon_0^2 A_e^2 U_0^2}{m^2 d_0^4}\,S_V(f)\,.
\label{eq:actuation_noise}
\end{equation}
Assuming low-noise electronics with an effective dynamic range of \(\sim190\,\mathrm{dB}\) over the \(0.1\)--\(10\,\mathrm{Hz}\) band, comparable to that achieved in state-of-the-art force-balance seismometers~\cite{Broadband_seismometer,garden_gate,LPS_new}, a \(1/f\) voltage-noise model gives
\[
\sqrt{S_V}(f) \simeq 3.3\times10^{-9}\times \left( \frac{1\,\mathrm{Hz}}{f}\right)^{1/2}\,\mathrm{V/\sqrt{Hz}}\,.
\]

The choice of the nominal electrode gap $d_0$ represents a design trade-off. Here we choose an intermediate gap that provides sufficient control authority while keeping the actuation noise subdominant. The maximum actuation acceleration then imposes a quasi-static leveling requirement along the sensitive \(x\)-axis
\begin{equation}
|\theta_{\rm lev}^x|\lesssim\frac{\epsilon_0 A_e U_0^2}{2m g_{\rm moon} d_0^2}
= 0.36\,\mu\mathrm{rad}\,.
\end{equation}
This requirement is about an order of magnitude more stringent than that of the LGWA design because of the reduced thermal-noise floor, calling for a dedicated high-stability leveling structure with active sensing and actuation. 
The orthogonal \(y\)-axis requirement is looser, as it mainly enters through a gravity-induced stiffness shift rather than direct gravity-to-readout coupling; a detailed estimate is given in Supplemental Material. 
A hierarchical leveling strategy is therefore required, particularly because the near-surface lunar regolith may be mechanically soft and the support points may creep after deployment. 
The station body is mechanically preleveled at the deployment site, an internal leveling table fine-levels the payload and suppresses residual DC and sub-millihertz tilt drift, and electrostatic control handles the remaining AC motion.

\textit{Temperature stability.---}
Diurnal temperature variations on the exposed lunar surface can reach several hundred kelvin. In a permanently shadowed region, however, the external environment is expected to be far more stable. The representative ambient temperature is around $40\,\mathrm{K}$ and a residual long-timescale fluctuation is of the scale of $\Delta T\sim 20\,\mathrm{K}$~\cite{PSR_temp,PSR_temp_simu}. For the TX seismometer, such a temperature environment is not favorable because fused-silica becomes more lossy at low temperature. A two-stage heating scheme is therefore envisioned: the interior of the seismic station is maintained near $200\,\mathrm{K}$, and the core TX payload is further heated and stabilized near $300\,\mathrm{K}$, where fused-silica fibers remain a favorable low-loss state.

Temperature fluctuations can couple into the TX readout through two main paths. The first arises through the control electronics. Assuming an optimized but realistic relative temperature coefficient $\beta = 2\,\mathrm{ppm/K}$, the corresponding temperature-induced voltage noise is
\begin{equation}
    S_V^{\rm temp}(f)=\beta^2 U_0^2 S_T^{\rm elec}(f)\,.  
\end{equation}
Where $S_T^{\rm elec}$ denotes the temperature noise spectrum of the electronics environment. This voltage noise then couples into the actuation force noise through Eq.~\eqref{eq:actuation_noise}.

A second coupling path is direct thermal expansion of the core payload, which changes the baseline between the interferometric sensor and the mirror on the test mass and thereby produces apparent displacement noise. For a lunar instrument, this structure is instead envisioned as a monolithic fused-silica breadboard with baseline $b=5\,\mathrm{cm}$ and thermal expansion coefficient $\alpha=5\times10^{-7}\,\mathrm{K^{-1}}$, giving an equivalent acceleration noise
\begin{equation}
S_{a,{\rm core}}^{\rm temp}(f)
=
32 \pi^4 \alpha^2 b^2 f^4 S_T^{\rm core}(f)\,.
\end{equation}
Where $S_T^{\rm core}$ denotes the temperature noise spectrum in the core payload region. Combining the electronics and baseline contributions, the total temperature-induced acceleration noise can be written as
\begin{equation}
S_a^{\rm temp}(f)
=
\frac{\beta^2\epsilon_0^2 A_e^2 U_0^4}{m^2 d_0^4}\,S_T^{\rm elec}(f)
+
S_{a,{\rm core}}^{\rm temp}(f)\,.
\label{eq:temp_noise}
\end{equation}

To set practical design targets for the TX seismic station, we assume the following temperature stability for the first- and second-stage temperature-control systems:
\[
\begin{aligned}
\sqrt{S_T^{\rm elec}}(f) &= 1.8\times10^{-2}
\left(\frac{1\,\mathrm{mHz}}{f}\right)^{1.3}
\,\mathrm{K/\sqrt{Hz}}\,,\\
\sqrt{S_T^{\rm core}}(f) &= 1.6\times10^{-7}
\left(\frac{1\,\mathrm{mHz}}{f}\right)^{3.6}
\,\mathrm{K/\sqrt{Hz}}\,.
\end{aligned}
\]
These levels correspond to a factor-of-10 relaxation of the in-flight temperature stability measured by LISA Pathfinder~\cite{LISA_temp_noise}. With these targets, the temperature induced noise is dominated by the electronics temperature path, while the core thermal-expansion term is suppressed by the tighter core-temperature stability and its \(f^2\) acceleration-referred coupling. The total temperature-induced noise therefore remains subdominant throughout the decihertz science band.

\textit{Summary and outlook.---}
In summary, the TX concept combines a fused-silica garden-gate suspension with an experimentally validated torsional readout architecture to address the low-frequency inertial-sensing requirements of lunar gravitational-wave detection. Future work should focus on a fully monolithic fused-silica implementation, which would suppress the clamping loss observed in the present prototype and potentially approach the material-loss limit. Further development of the supporting subsystems is also essential, as temperature control, precision leveling, and low-noise electrostatic actuation are common challenges for high-precision lunar seismometer concepts. Integrating these subsystems at the required noise levels remains a key step toward a realistic lunar instrument.

\section*{Acknowledgment}

Y. X. and H. M. would like to acknowledge the support from the ``Gravitational Wave Detection'' program (2023YFC2205800) and ``JinPing frontier research'' program (12441503) funded by the National Natural Science Foundation of China. D. M. would like to acknowledge the support of the Institute for Gravitational Wave Astronomy at the University of Birmingham and UKRI “The next-generation gravitational-wave observatory network” project (Grant No. ST/Y00423X/1). H. Y. is supported by the Natural Science Foundation of China (Grant 12573048).

\section*{Data Availability}
The data and analysis scripts supporting the findings of this article are openly available~\cite{TX_Data}.
\bibliography{TX_reference}

\end{document}